\documentclass[preprint]{aastex}

\newcommand{\dif}{d}
\newcommand{\sect}{\S}
\newcommand{\ee}[1]{\times10^{#1}}
\newcommand{\ud}[2]{$^{+#1}_{-#2}$}
\newcommand{\cenx}{\objectname[]{Cen~X-3}}
\newcommand{\cenxsp}{\objectname[]{Cen~X-3 }}
\newcommand{\asca}{{\it ASCA\/} }
\newcommand{\ergs}{erg\,s$^{-1}$}
\newcommand{\kms}{km\,s$^{-1}$}
\newcommand{\vovf}{\frac{v_0}{v_\infty}}
\newcommand{\lvom}{$Lv_\infty/\dot{M}$}
\newcommand{\movf}{$\dot{M}/v_\infty$}

\shorttitle{The X-Ray Photoionized Wind in
 \cenx/\objectname[]{V779~Cen } }
\shortauthors{Wojdowski, Liedahl, \& Sako}

\begin{document}

\title{The X-Ray Photoionized Wind in
\cenx/\objectname[]{V779 Cen}}

\author{Patrick S. Wojdowski, Duane A. Liedahl}
\affil{Physics Department, Lawrence Livermore National Laboratory}
\affil{P.O. Box 808, Livermore, CA 94551}
\email{patrickw@virgo.llnl.gov}
\email{duane@virgo.llnl.gov}
\and
\author{Masao Sako}
\affil{Columbia Astrophysics Laboratory and Department of Physics,
Columbia University, 538 West 120th St., New York, NY 10027}
\email{masao@astro.columbia.edu}

\begin{abstract}
We analyze the {\it ASCA\/} spectrum of the \cenxsp X-ray binary system in
eclipse using atomic models appropriate to recombination-dominated
level population kinetics in an overionized plasma.  In order to
estimate the wind characteristics, we first fit the eclipse spectrum
to a single-zone photoionized plasma model.
We then fit spectra from a range of orbital phases using global models
of photoionized winds from the companion star and the accretion disk
that account for the continuous distribution of density and ionization
state.  We find that the
spectrum can be reproduced by a density distribution of the form
derived by \citet*{cas75} for radiation-driven winds with
$\dot{M}/v_\infty$ consistent with values for isolated stars of the
same stellar type.  This is surprising because the neutron star is
very luminous ($\sim10^{38}$\,\ergs) and the X-rays from the neutron
star should ionize the wind and destroy the ions that provide the
opacity for the radiation-driven wind.
Using the same functional form for the density profile, we also fit
the spectrum to a spherically symmetric wind centered on the neutron
star, a configuration chosen to represent a disk wind.  We argue that
the relatively modest orbital variation of the discrete spectrum rules out a
disk wind hypothesis. 
\end{abstract}

\section{Introduction}
\label{intro}

Since the early observations of the High Mass X-ray Binary (HMXB)
Cen~X-3, it has 
been known that the system exhibits a residual X-ray flux in eclipse,
indicating that the X-rays are scattered or otherwise reprocessed, as
might occur in a wind \citep{sch72}.  A similar phenomenon has been
observed in other HMXBs --- for example,
\citet{bec78} observed a residual 
eclipse flux from Vela~X-1.  It has been realized since the far
ultraviolet became accessible with rocket and satellite-borne
instruments that isolated hot stars expel material at
rates of order $10^{-6}\,M_\sun$\,yr$^{-1}$ \citep{mor67}, and so the
existence of winds in HMXBs might have been expected.  It was
shown that in isolated hot stars, strong winds are driven by
transfer of the outward momentum of the ultraviolet stellar radiation
to the matter through line absorption \citep{luc70,cas75}. 

However, it was pointed out by \citet{hat77} that in an X-ray
binary, X-rays from the compact object would ionize a part of the
wind around the X-ray source to such a high degree that the ions
with line transitions which enable radiative driving would
not be present.  Presumably, if an X-ray source were luminous enough, it
could completely shut off the radiation driven wind on the X-ray
illuminated side of the companion star.  Calculations by
\citet{mgr82} showed that radiative driving is disabled for X-ray
luminosities larger than $\sim5\ee{34}$\,\ergs.  As HMXBs
typically have luminosities in the range $10^{36}$--$10^{38}$\,\ergs,
this would indicate that radiative driving could not function at all
in X-ray binaries.  However, this calculation assumed that the wind is
optically thin to the ionizing X-rays.  The effects of the wind's
opacity to X-rays has been explored by \citet{mas84} and by
\citet{ste91}.  These calculations considered the formation of a
He$^{++}$/He$^+$ Str\"omgren surface in the wind.  On the He$^+$ side
of this boundary, all elements have lower ionization, and 
more of the ions necessary for radiative driving are present.
\citet{mas84} concluded that most wind-fed X-ray binaries must contain 
Str\"omgren boundaries and hence radiative winds.  This conclusion is
limited to wind-fed systems because of the coupling of the wind
parameters to the accretion luminosity in these systems.  An
increase in the luminosity of the 
X-ray source tends to increase the volume of the He$^{++}$ region,
however, to increase the luminosity it is necessary to increase the
mass loss rate of the companion star or decrease the wind velocity,
both of which tend to decrease the volume of the He$^{++}$ region.
The investigation of \citet{ste91} was not limited to wind-fed
systems.  It showed that X-ray luminosities as large as
$\sim10^{36}$\,\ergs ~diminished the wind velocity and mass loss rate
but did not completely shut off the wind.  However, for luminosities
larger than 
$4\ee{36}$\,\ergs ~\citet{ste91} was ``unable to to find dynamical
solutions'' for the wind.  \cite{ho87}  explored the behavior of high
luminosity (exclusively) wind-fed X-ray binaries.  However, they did
not explicitly consider optical depth effects and their condition for
X-ray ionization turning off the wind is questionable, even for the
optically thin 
case \citep[c.f.\ ][]{ste91}. We regard the nature of
winds in X-ray binaries outside of the X-ray shadow of the companion
as unresolved, and especially problematic in the case of high
luminosity ($L_{\rm X}\sim10^{38}$\,\ergs) systems.

A wind driving mechanism which does not depend on the presence of ions
with UV resonance lines is thermal pressure due to X-ray heating of the exposed
face of the companion star (also referred to as evaporative winds).  This
mechanism was invoked by \citet{bas73} and by \citet{aro73} to explain
the mass transfer in \objectname[]{Her~X-1}, though \citet{alm74} and
\citet{mcr75} found that a thermal wind alone could not power the
X-ray source.  
\citet{day93a} showed that an X-ray excited wind could
account for the mass transfer in \cenx, as well as explain the
extended eclipse transitions seen in the source.  At least one
simulation of the disk-fed high-mass X-ray binary
\objectname[]{LMC~X-4}, in which the effects of X-ray heating were
included, showed that a stronger wind was driven from the accretion
disk, and that the structure of the wind was dominated by the disk
wind \citep{owe97}.

Before the launch of {\it ASCA\/} in 1993 the energy resolution of
most cosmic X-ray detectors was rather poor ($\Delta E/E\sim$10--20\%)
and therefore, these instruments were unable to detect the X-ray
emission lines which are a signature of optically thin, highly ionized
gas.  Nor is there any other waveband in which discrete emission from
very highly ionized gas could be detected.  Most observational X-ray
examinations of winds in HMXBs focused on absorption in low-ionization
material, iron fluorescence, which is produced primarily in
low-ionization material, and the Compton scattering continuum which is
insensitive to the ionization state.
\citep{kal82b,nag86,sat86,cla88,hab89,woo95}.
The lack of a regular low-ionization wind on the X-ray illuminated
side of HMXBs, and therefore the presence of a
high-ionization wind there, has been inferred from decreases in
equivalent widths and velocities of the P~Cygni profiles of
ultraviolet resonance lines away from X-ray eclipse
\citep{dup80,vdk82,ham84,vrt97}. 
However, direct study
of the highly ionized material requires X-ray spectroscopy.  The
Solid-state Imaging Spectrometers \citep[SIS, ][]{gen95} on board
\asca have energy resolution of a few per cent which allows
identification and study of many previously undetectable X-ray
spectral features.  With the SIS detectors, recombination and
fluorescence emission features were seen from several ions from the
HMXBs Vela~X-1 \citep{nag94}, Cen~X-3 \citep{ebi96}, and Cyg~X-3
\citep{kit94,lie96,kaw96}.

When a compact X-ray source is occulted by its companion, the emission
spectrum from an extended wind can be studied without the confusion
from the more intense, generally featureless, spectrum of X-rays from
the neutron star.  \citet{sak99} studied the eclipse spectrum of the
low luminosity HMXB Vela~X-1 obtained by \asca and estimated the rate
of mass loss.  Though a highly ionized wind exists in Vela~X-1, they
found that that most of the mass is inside dense clumps, which are not
highly ionized.  This allows for the possibility that radiation
imparts its outward momentum to the clumps which then drag the hot,
diffuse wind outward.  Presumably, these clumps could be destroyed (or
inhibited from forming) by a more luminous X-ray source such as
Cen~X-3 or SMC~X-1.  In fact, \citet{ebi96} showed that the equivalent
width of the 6.4\,keV iron fluorescence line in Cen~X-3 was nearly
constant with orbital phase, indicating that most of the
low-ionization material in the system is located near the neutron
star, and that very little low-ionization material is found in the
extended wind.  This is also confirmed by pulsations in the 6.4\,keV
line \citep{day93b,aud98}.  \citet*{woj00} used hydrodynamic
simulations of the wind in the most luminous persistent HMXB SMC~X-1
by \citet{blo95} to predict the X-ray eclipse spectrum of that system
and compared it to a spectrum obtained with {\it ASCA\/}.  In the
simulation, a tenuous, very highly ionized wind formed on the X-ray
illuminated side and a denser wind developed on the X-ray shadowed
side.  However, dense finger-like structures protruding from the
shadowed side of the companion were swept into the X-ray illuminated
region by the Coriolis force.  The calculations showed that the
reprocessed radiation from the tenuous gas was dominated by Compton
scattering and the denser gas emitted copious amounts of recombination
radiation.  The observed spectrum was nearly featureless however, and
\citet{woj00} concluded that the dense fingers that appeared in the
simulation could not be present in the wind of SMC~X-1.

Cen~X-3 is one the most luminous, persistent known HMXBs in the
Galaxy.  It consists of a 4.8 second pulsar in a 2.08 day eclipsing
orbit \citep{sch72} with its O6--8 III type \citep{con78,hut79}
companion V779 Cen.  The high X-ray luminosity of Cen~X-3 makes it an
excellent candidate for the study of X-ray photoionized winds.  It was
observed by {\it ASCA\/} over approximately half an orbit, which
included an eclipse.  \citet{ebi96} found several emission lines in
this data set, which were mostly from hydrogen-like ions.  From the
intensities of these lines, they made estimates of the scale and
physical conditions of the wind.

We re-analyze the data set of \citet{ebi96} using the observed spectra
to test physically motivated wind models with the goal of providing
constraints on the wind driving mechanism.  In \sect~\ref{reduction}
we describe our reduction of the data.  In \sect~\ref{single-zone}, we
calculate emission spectra for photoionized plasmas using a list of
$\sim$3000 lines and emission features, fit the observed spectra using
single zone emission spectra and then use the results to estimate wind
parameters.  In \sect~\ref{global_models} we calculate spectra using
explicit parameterized models of the wind density distribution, and
fit the observed spectra to determine wind parameters more accurately.
We test two explicit global wind models: 1) a stellar wind from the
companion with the velocity profile of a radiatively driven wind and
2) motivated by the accretion disk wind in the simulation of
\citet{owe97}, a wind with the same velocity profile but centered on
the neutron star.  In \sect~\ref{optdep} we justify an assumption that
the wind is optically thin to X-rays which is used in previous
sections.  In \sect~\ref{discussion}, we discuss the implications of
our results.

\section{Data Reduction}
\label{reduction}

We obtained the screened REV2 {\it ASCA} event data from the 1993 June
24-25 observation of Cen~X-3 across an eclipse from the HEASARC
archive.  All manipulation of the data was done with FTOOLS v4.2
\citep{ft42} and all of the programs mentioned in this section are
from that package.  We divided the data into the same four time
segments as \citet{ebi96}, corresponding to
phase ranges $-0.31$ to $-0.29$, $-0.23$ to $-0.08$, $-0.08$ to
$0.13$, and 0.14 to 0.20.  The data was taken in a mix of FAINT and
BRIGHT modes.  For the data taken in FAINT mode, we used the files
which had been converted to BRIGHT mode on the ground in order to have
homogeneous data.  For each of the time segments, we extracted all
counts from inside a circle of radius 191\arcsec.  We extracted a
background spectrum from an annulus of inner radius 191\arcsec~ and
outer radius 382\arcsec.  During the observation, the center of the
image was placed so that the source counts were distributed over all
four of the detector chips.  During the eclipse phase, all four chips
were on but during the rest of the observation only one of the four
chips was on, resulting in a $\sim$45\% collection efficiency due to
the placement of the source near the chip boundary.  The spectra were
extracted with the standard channel binning applied by XSELECT
resulting in 512 energy channels.  The spectra from the different
chips of each detector and from the two SIS detectors from the same
time interval were added using ADDASCASPEC.  The energy channels in
the range 2.9--8.0\,keV were binned additionally by a factor of two.
Additional binning was applied to the energy channels in the range
8.0--10.0\,keV so that each channel in the eclipse spectrum had at
least 50 counts.

The analysis tools we used to compute the detector response include
the reduction in effective area due to the fact the source areas we
have chosen do not include all of the photons focussed by the mirrors.
However, the background regions also contain some source photons which
are subtracted from the source spectrum in computing the background
selection region. 
Our source region (for all chips on) contains approximately 80\% of
the photons for a point source at the center and our background region
contains approximately 15\% \citep{ser95}.  The background spectrum is
multiplied by the area of the source region and divided by the area of
the background region and then subtracted from the source spectrum.
Because the background region is larger than the source region by a
factor of 3, approximately 5\% of the source flux is subtracted from
the source spectrum.   This effect is not accounted for by our
analysis.  This problem is compounded by the fact that the
image of Cen~X-3 is further smeared, in a way which depends on energy,
due to an X-ray halo and this halo is delayed in time relative to the
direct photons.  However, the radius of the halo is approximately the
size of the source region and the surface brightness due to the halo
is generally no larger than that due to the image of the direct
photons \citep{woo94} and therefore leads to a similar subtraction of
source photons.  Because of the complexity of these effects, we do not
try to account for them explicitly but note that fluxes (and
quantities proportional to flux) which we derive are too small by
approximately a factor 10\%.  This error does not qualitatively affect
any of the results we derive in this work.

\section{Single-Zone Spectral Models}
\label{single-zone}

Emission spectra from X-ray binaries are generally interpreted with
the assumption that photoionization from the compact X-ray source is
the dominant source of ionization in the circumstellar plasma.  This
is justified by the relative values of the luminosities of the X-ray
sources and the densities and linear scales of the system.  While it
is straightforward to measure X-ray luminosities and, where orbital
parameters are available from pulsar timing and optical spectroscopy,
linear scales in X-ray binaries, determinations of the matter density
that do not depend on the assumption of photoionization equilibrium
are less reliable.  Photoionization equilibrium has been inferred
directly from observed X-ray recombination spectra \citep[e.g.,
][]{lie96}.  In the case of Cen~X-3 however, there are no obvious
spectral signatures of recombination dominance.  The observed spectrum
consists mainly of Ly$\alpha$ transitions from hydrogen-like ions
which, in principle, can be produced in plasmas where collisions
dominate the ionization.  We therefore attempt to fit the eclipse
spectrum of Cen~X-3 with spectral models of emission from
collisionally ionized (coronal) plasma as well as photoionized
plasmas.

The X-ray spectrum observed from Cen~X-3 includes at least the
following components: direct emission from the neutron star, X-rays
from the neutron star which have been scattered in the wind, continuum
and line emission from the wind and other circumstellar material, and
X-rays scattered from interstellar dust grains.  All of these
components contribute to the observed continuum, but only the wind and
circumstellar material can emit lines.  We found that, in general, it
was possible to fit the continuum using two power laws, each absorbed
by a different column density.  The more highly absorbed power-law may
correspond approximately to the X-rays from the neutron star viewed
through some dense component of circumstellar material and the less
absorbed power-law to the neutron star continuum scattered in the
extended wind and by dust.  However, since our primary goal is to
extract and interpret the emission line spectrum, we do not attempt to
constrain the parameters of these power laws in a manner that would
force them to correspond to physical sources of continuum emission
\citep[c.f., ][]{ebi96}.  We interpret the lines, except for the
6.4\,keV Fe K$\alpha$ line as emission from the extended wind.  These
plasma emission models are described in detail below.  The spectral
model is,
\begin{equation}
{\cal F}(\epsilon) = \mathrm{e}^{-\sigma(\epsilon)N_{\rm H1}} 
\left\{f_{\rm pl1}(\epsilon)+f_{\rm plasma}(\epsilon) + \\ 
\mathrm{e}^{-\sigma(\epsilon)N_{\rm H2}}
[I_{\rm line}\delta(\epsilon-\epsilon_{\rm line}) + 
f_{{\rm pl}2}(\epsilon)]\right\}
\label{spec_exp}
\end{equation}
where the power laws,
\begin{equation}
f_{{\rm pl}i}(\epsilon)=K_{{\rm pl}i}\left(\frac{\epsilon}{1\,{\rm
keV}}\right)^{-\alpha_i},
\end{equation}
$f_{\rm plasma}$ is the plasma emission model, $\sigma(\epsilon)$ is
the interstellar absorption cross-section of \citet{mor83}, $N_{{\rm
H}i}$ are the absorption column densities, $I_{\rm line}$ is the line
photon flux and $\delta$ is the Dirac delta function, used for the the
6.4\,keV Fe K$\alpha$ line complex. The energy of the Fe K$\alpha$
line, $\epsilon_{\rm line}$, was constrained to be in the range
6.3--6.5\,keV. The Fe K$\alpha$ line was the only fluorescent feature
required to fit the data in any of our fits.  \citet{ebi96} identified
an emission feature at energy 1.25$\pm$0.04\,keV with the 1.25\,keV
\ion{Mg}{1} K$\alpha$ fluorescent line.  However, in all of our
spectral models described here, this emission feature is fully
accounted for by \ion{Ne}{10} Ly$\beta$ (1.21\,keV).

When the neutron star is eclipsed by the companion star, the X-rays
from the extended wind only are observed.  Therefore, for the purpose
of testing basic plasma emission models, we fit only the spectrum from
eclipse.  For our spectral fits, we used the XSPEC spectral fitting
program \citep[v10.0, ][]{arn96}, importing our own models for
emission from photoionized plasmas.

\subsection{Collisional Ionization Equilibrium}
\label{cie}

We fit the eclipse spectrum, using for $f_{\rm plasma}$ the emission
spectrum of an isothermal plasma in collisional ionization equilibrium
(CIE).  Also referred to as coronal equilibrium, this describes a
situation where recombination is balanced by collisional ionization by
electrons, and ionization by radiation is negligible.  We use the
MEKAL model \citep{mew95} contained in XSPEC for the CIE plasma
emission.  The three parameters of the MEKAL model are the
temperature, metal abundance, and the normalization.  The emission
processes in CIE are determined by two-body interaction rates, so the
normalization of the flux is proportional to the emission measure
($E\equiv\int n_{\rm e}^2\dif V$) divided by the square of the
distance to the source.  The best fit parameters are shown in
Table~\ref{io_mod_fit}, and the spectrum is plotted in
Figure~\ref{ecl_mekal}.

The MEKAL model, with best-fit parameters, which reproduces most of
the the observed emission lines, also accounts for all of the
continuum emission below $\sim$4\,keV through bremsstrahlung radiation
from the same gas.  For
parameters for which the MEKAL model accounts for the observed line
emission, the model also accounts for all of the continuum emission
below $\sim$4\,keV.  However, collisionally ionized gas must also
scatter X-rays from the neutron star.  In the \asca band,
optically thin electron scattering reproduces the continuum shape
of the X-ray source with a fractional luminosity relative to the
source spectrum equal to:
\begin{equation}
\frac{L_{\rm scat}}{L}=\frac{\sigma_{\rm T}}{4\pi}\int\frac{n_{\rm e}}{r^2}\dif V,
\end{equation}
where $\sigma_{\rm T}$ is the Thomson cross-section and $r$ is the
distance from the compact source.  We can estimate this fraction by
taking the orbital separation $a$ as the linear scale of the system
and setting:
\begin{eqnarray}
r  = a, & n_{\rm e}=(E/V)^{1/2}, & V =\frac{4\pi}{3}a^3 . 
\label{fiduc_dims}
\end{eqnarray}
Then,
\begin{equation}
\label{eq:scat}
\frac{L_{\rm scat}}{L}\approx(12\pi)^{-1/2}\sigma_{\rm T}
E^{1/2}a^{-1/2}
=1.6\ee{-2}
\end{equation}
If the neutron star has an intrinsic spectrum in the 2--10\,keV band
which is a power law with photon index 1.5, and luminosity
$\sim10^{38}$\,\ergs, then, using the system parameters of
Table~\ref{syspars}, the Compton scattered luminosity in the
2--10\,keV band should be $\sim10^{36}$\,\ergs.  This corresponds to a
power-law normalization of $I_{\rm
pl}=1.6\ee{-2}$\,s$^{-1}$\,cm$^{-2}$\,keV$^{-1}$.  However, the upper
limit on $I_{\rm pl1}$ from the fit,
$1.1\ee{-4}$\,s$^{-1}$\,cm$^{-2}$\,keV$^{-1}$, corresponds to a
scattered luminosity of $7\ee{33}$\,\ergs.  In principle, it is
possible that the Compton scattered continuum could be reduced
relative to the thermal emission due to clumping of the wind.  If the
wind is clumped with volume filling factor $f$, then the density would
have to be increased by a factor $f^{-1/2}$ to preserve the emission
measure.  The new Compton scattered flux would then be changed by a
factor $f^{1/2}$ since its magnitude is proportional to $nV$.  A very
small filling factor ($f\approx10^{-4}$) would be necessary however
and we regard this as unlikely.  We therefore reject this model.

Increasing the metal abundance in the gas increases the flux of the
lines relative to the bremsstrahlung continuum, thereby providing
``room'' for a scattered continuum component.  Therefore, we tried
another fit in which we allowed the metal abundance to be free.  For
this fit, we tied the normalization of the first power law (pl1) to
the emission measure of the MEKAL component as described above
(Eq.~\ref{eq:scat}).  XSPEC does not allow the user to set one
parameter to the square root of another parameter, so we fixed the
power-law normalization according to a trial value of the emission
measure, fit, and iterated to find a best fit emission measure with a
consistent power-law normalization.  This procedure resulted in a
statistically acceptable fit (Table~\ref{io_mod_fit},
Figure~\ref{ecl_mekalv}).  The lower limit on the abundance derived
from this procedure, however, is 25 times solar, which we believe to
be implausibly high. and so we reject this model as well.

\subsection{Photoionization Equilibrium}
\label{photoio}

Having rejected plausible collisional ionization spectral models, we
proceed to fit the observed spectra using model emission spectra from
photoionized plasmas.  The radiation spectrum due to recombination
consists of radiative recombination continua (RRC) from free electrons
recombining directly to bound states and lines which are produced
through radiative cascades.  The volume emissivity of the
recombination feature $k$ is $j_k= n_{\rm e} n_{i+1} \alpha_k(T)$
where $n_{i+1}$ is the density of the recombining ion to which the
electron recombines.  The function $\alpha_k(T)$ is an effective
recombination rate coefficient (the recombination coefficient for
recombinations that produce transition $k$) that depends on atomic
parameters and the level population kinetics.  In a gas which is
optically thin and has density low enough such that all ions may be
assumed to be in the ground state (i.e., the density is low compared
to the critical density for all important transitions), the ion
fractions and temperature are, for a given spectrum of ionizing
radiation, functions only of the ionization parameter
\citep*[$\xi=L/nr^2$, ][]{tar69}.  Therefore, for a photoionized
plasma, the emissivity of any feature may be written $j_k=n_{\rm
e}^2f_k(\xi)$, and, for the entire spectrum, $j_\nu/n_{\rm e}^2$ is a
function only of $\xi$.

To calculate the temperature and ion fractions as a function of $\xi$,
we need to know the X-ray spectrum of the neutron star.  The best way
to do this is to observe the X-ray spectrum from the neutron star
directly while outside of eclipse.  This {\it ASCA} data set includes
data outside of eclipse.  However, the observed spectra in these
phases outside of eclipse differs greatly from the intrinsic spectrum
of the neutron star because at the time of the observation the line of
sight to the neutron star contained circumstellar material that
substantially absorbed the direct X-rays.  When circumstellar material
passes in front of the neutron star, the direct X-rays are absorbed
preferentially at low energies.  Even when direct low energy X-rays
are absorbed completely, low energy X-rays are still observed from
scattering and emission from the extended wind, just as during eclipse
by the companion star.  The residual X-rays from scattering are
generally not pulsed.  The X-rays from Cen~X-3 in this data set show
both a spectrum which is deficient in low-energy X-rays compared to
other observations \citep{nag92,san98} and also are not pulsed at low
energy \citep{ebi96}.  \citet{nag92} used similar evidence to
demonstrate that a ``pre-eclipse dip'' observed by {\it Ginga} was due
to absorption.  In these out-of-eclipse intervals, the large
absorption of the direct X-rays makes the strength of the direct flux
comparable to and difficult to distinguish from the residual flux from
scattering and wind emission, and it is impossible to isolate the flux
from the neutron star in the observed spectrum.  For the intrinsic
spectrum of the neutron star, we instead use a spectrum obtained by
\citet{san98} with {\it BeppoSAX}.  \citet{bur00} have shown that in
this observation, the X-ray flux is pulsed down to the energy band
below 1.8\,keV indicating that it originates in the proximity of the
neutron star.  There is an additional advantage to using {\it
BeppoSAX} in that it can measure the spectrum up to $\sim 200$\,keV.
For $\xi\gtrsim 10^4$, the ions are nearly fully stripped, and the
temperature is determined by Comptonization.  The Compton temperature
is highly sensitive to the presence of high energy photons.  We used
the ``Lorenzian'' model of \citet{san98}, extending the power-law
continuum to low energy by setting the absorption column to zero, and
excluding the cyclotron absorption feature, since a value for the
depth of this feature is not provided.

We used the XSTAR program \citep[v1.43, ][]{kal99} to compute ion
fractions and temperatures for 100 uniformly spaced values of
$\log\xi$ from $-2$ to 6 using the \citet{san98} spectrum.  In the
XSTAR calculation, we used a luminosity of 10$^{30}$\,\ergs ~and a
constant density 10$^{-2}$\,cm$^{-3}$ --- again, for an optically thin
gas, the absolute values of the luminosity and the density are not
important.  We then calculated the recombination spectrum at each
value of $\xi$ using the ion fractions and temperatures from XSTAR,
and line and RRC powers calculated using the Hebrew
University/Lawrence Livermore Atomic Code \citep[HULLAC, ][]{kla77}
and the Photoionization Cross-Section code \citep*[PIC, ][]{sal88} for
recombination cross-sections.  These atomic models have been used to
analyze the recombination spectrum of Cyg~X-3 \citep{lie96} and
Vela~X-1 \citep{sak99} and are described further there.

We fit the eclipse spectrum using for $f_{\rm plasma}$ our
recombination model spectra plus a corresponding bremsstrahlung
component.  The bremsstrahlung component is the BREMSS model from
XSPEC.  We tied the normalization of the bremsstrahlung component to
that of the recombination component such that the emission measures
would be equal.  We set the temperature of the bremsstrahlung model to
be a locally linear approximation to the function $T(\log\xi)$
computed with XSTAR in the neighborhood of the best fit value of
$\log\xi$.  The best fit spectrum is shown in Figure~\ref{ecl_xi}, and
the best fit parameters are shown in Table~\ref{io_mod_fit}.  The best
fit value of the ionization parameter ($\log\xi=3.19$) corresponds to
an electron temperature of 0.41\,keV.  At this temperature, the
radiative recombination continua are broad, and so it is not
surprising that they do not appear prominently in the spectrum.  It is
because of this that \citet{ebi96} were able to fit the spectrum using
only lines and not RRC.  This demonstrates that though the presence of
prominent narrow RRC in emission spectra indicates photoionization
\citep{lie96}, the lack of narrow RRC does not necessarily indicate
some other type of equilibrium.  It can be seen that the
bremsstrahlung component is relatively faint, thus allowing a
power-law continuum at low energies.  The luminosity implied for the
power-law component \#1 is $4\ee{35}$\,\ergs which is much closer to
the expected Compton scattered luminosity of $\sim10^{36}$\,\ergs.  A
quasi-continuum of blended lines and RRC accounts for 21\% of the
photon flux in the 1--3\,keV band.  This differs significantly from
the individual line fits of \citet{ebi96} in which the lines
constitute only 5\% of the flux in this band.

From this single-zone fit, it is possible to make a crude estimate of
the X-ray luminosity and parameters of the wind in Cen~X-3, if we take
the orbital 
separation to be the characteristic linear scale of the system, and
use our best fit value of the emission measure as in
Equation~\ref{fiduc_dims}.  We then we find for a
characteristic wind density
$n\sim5\ee{10}\,{\rm cm}^{-3}(d/10\,{\rm kpc})^{-1}$.  From the
definition of $\xi$, we then have
\begin{equation}
L=\xi nr^2\sim 1.4\ee{38}\,{\rm erg\,s^{-1}}(d/10\,{\rm kpc})^{-1}.
\end{equation}
Note that this value of the luminosity is 
derived only from the flux of the recombination spectral features and
the known system dimensions, yet it comes very close to the luminosity
derived for Cen~X-3 by measuring the continuum directly during the
high state \citep{nag92,san98}. 
\citet{ebi96} derived similar, though somewhat larger, estimates for
the characteristic density and luminosity
($1.6\ee{11}$\,cm$^{-3}$ and $1.8\ee{38}$\,\ergs ~for $d$=10\,kpc)
using an emission 
measure derived from the fluxes of the 6.7\,keV and 6.9\,keV lines of,
respectively, helium-like and hydrogen-like iron in the eclipse
and egress spectra and by assuming a value of $\log\xi=3.4$ from their
ratio.  Our values differ because we use
slightly different values for the linear scale and also because
our values of $\xi$ and $E$ are derived from fitting all of the lines
and not just the hydrogen- and helium-like iron lines.  

We now estimate the characteristics of the stellar wind from this
derived value of the density.  In a spherically symmetric steady-state
wind, continuity demands that:
\begin{equation}
\frac{\dot{M}}{v(R)}={4\pi R^2 \mu n(R)}
\label{mvfr}
\end{equation}
where $\dot{M}$ is the mass-loss rate, $v$ is the velocity, $R$ is the
distance from the center of the star, $n$ is density of hydrogen
(neutral and ionized), and the quantity $\mu$ is the gas mass per
hydrogen atom which we take to be 1.4$m_{\rm H}$.  Again using the
fiducial dimensions,
$\dot{M}/v\sim4\ee{-9}M_\sun$\,yr$^{-1}$\,(km\,s$^{-1}$)$^{-1}$($d$/10\,kpc)$^{-1}$
which, as will be discussed in \sect~\ref{discussion} below, is near
values for isolated stars with spectral type similar to V779~Cen.
While this model provides a statistically acceptable fit to the data,
it is unlikely that the entire emission line region can be
characterized by a single value of $\xi$.  Therefore, we suggest that
this model provides no more than a semi-quantitative description of
the data.  In the next section we use explicit wind models and spectra
of all of the full phases of the observation to derive these wind
parameters more accurately and to explore the wind geometry.

\section{Global Photoionized Wind Models}
\label{global_models}

We use the recombination spectra described in the previous section to
derive spectra for explicit wind density distributions.  In order to
calculate the global recombination spectrum of photoionized gas, we
use the differential emission measure formalism \citep[DEM,][]{sak99}.
For an arbitrary distribution of gas, the total spectrum of
recombination radiation is given by
\begin{equation}
\label{spec_int}
L_\nu=\int \frac{j_\nu(\xi)}{n_{\rm e}^2}\left[\frac{\dif E}{\dif
\log\xi}\right]\dif\log\xi ,
\end{equation}
where the quantity in brackets is the DEM
distribution, which hereafter will be referred to as $D(\xi)$.  In
practice, we integrate only the volume of gas that is visible to the
observer and refer to this as the apparent DEM.  To
calculate the DEM distribution, it is necessary to know only the
density distribution of the gas and the luminosity of the radiation
source.  To calculate $j_\nu(\xi)/n_{\rm e}^2$, it is necessary to know only
the spectral shape of the ionizing radiation and the elemental
composition of the gas.  In this work, we calculate the spectrum of
recombination emission from the wind of Cen~X-3 for different models
of the wind using different luminosities for the 
neutron star, but always using the same spectral shape for the neutron
star X-ray emission and always assuming solar abundances
\citep{and89}.  The DEM formalism requires us to evaluate 
$j_\nu(\xi)/n_{\rm e}^2$ only once for a number of values $\xi$, and then,
for different matter distributions and luminosities, to calculate the
DEM distribution and evaluate the integral in Equation~\ref{spec_int}.

To calculate the DEM distribution for a given density
distribution and luminosity, we divide the binary system into
spatial cells.  We then calculate $\xi$
and the emission measure for each spatial cell.  The emission measure for
each cell is then added to a running sum of the emission measure for
a $\xi$ bin.  Dividing the emission measure in each
bin by the width of that bin in $\log\xi$ gives the differential
emission measure.  If the density distribution is described by
parameters such that the density everywhere scales linearly with one
parameter, which we will call $\eta$, then to calculate the DEM
distribution for new values of $\eta$ and the luminosity $L$ it is not
necessary to recalculate the DEM distribution by summation over the
spatial cells.  For new values $L^\prime$ and $\eta^\prime$, the
new DEM distribution $D^\prime$ is related to the old DEM distribution
$D$ by
\begin{equation}
\label{dem_scal}
D^\prime\left(\frac{L^\prime \eta}{L \eta^\prime}\xi\right)=
\left(\frac{\eta^\prime}{\eta}\right)^2 
D(\xi).
\end{equation}
This identity is derived in Appendix~\ref{app:dem}.  For a change in
the luminosity and density parameter such that
$L^\prime/\eta^\prime=L/\eta$, the DEM distribution is changed only by
a constant factor $(\eta^\prime/\eta)^2=(L^\prime/L)^2$ and therefore
so is the total emission spectrum.

\subsection{DEM Distributions for Model Winds}
\label{dem_mods}

As alluded to in \sect~\ref{intro}, the structure of stellar winds in
X-ray binaries are likely to be rather complicated.  In addition to
the effects of X-ray photoionization on the radiation driving, the
gravity of the compact object and orbital motion may have a
significant effect on the wind density distribution \citep[e.g.,
][]{fri82}.  Though numerical simulations may account for many of
these effects, we choose instead to use a simple, spherically
symmetric density distribution with free parameters such that the
density is easily recalculated for a change in parameters.  We
approximated the geometry of the X-ray binary system as a spherical
star in orbit with a point-like neutron star emitting X-rays
isotropically.  For the dimensions of the system, we assumed the
values in Table~\ref{syspars}.  We used density distributions for
spherically symmetric, radiation driven winds \citep{cas75,kud89} from
the companion star.

Modeling accretion disk winds introduces a new set of
complications.  They may
be driven thermally \citep[e.g.,][]{woo96}, radiatively
\citep*[e.g.,][]{pro99}, or flung out along rotating magnetic field
lines \citep[magnetohydrodynamically, e.g.,][]{bla82,pro00}.  While
stellar winds may be symmetrical in two dimensions (azimuth and
altitude), disk winds are necessarily symmetric in no more than one
dimension (azimuth).  However, in order to derive some observational
constraints on a disk wind hypothesis, we use the same spherically symmetric
symmetric radiation-driven velocity profile as for the stellar wind
but center the wind on the neutron star.  

The explicit form of the wind velocity profile is given by:
\begin{equation}
\label{vofr}
v(R)=v_0 + (v_\infty-v_0)(1-R_{\rm in}/R)^\beta
\end{equation}
where $v_0$ and $v_\infty$ are the wind velocities at the stellar
surface and at infinity and the parameter $\beta$ describes the
acceleration of the wind.  With $\beta=0$, this equation describes a
wind which is immediately accelerated to its terminal velocity.  The
variable $R$ is the distance from the center of the wind (the center
of the companion for the stellar wind or the neutron star for the disk
wind).  For the stellar wind, the wind
begins at the surface of the companion so $R_{\rm in}=R_\star$.  For a
disk wind however, no such natural inner radius exists and therefore,
for the disk wind, we make $R_{\rm in}$ an extra parameter of the
model and assume that the volume for which $R<R_{\rm in}$ is empty.
The value of $R_{\rm in}$ may correspond, for example, to a
characteristic radius on the accretion disk from which the wind
arises.  Because an accretion disk may be no larger than the Roche
lobe of the accreting object, the Roche lobe radius should provide an
approximate upper limit on $R_{\rm in}$.

A rearrangement of Equation~\ref{mvfr} gives the density distribution.
\begin{equation}
n(R)=\frac{\dot{M}}{4\pi R^2 \mu v(R)}.
\end{equation}
In order to make use of the scaling relation of
Equation~\ref{dem_scal}, we include the velocity profile explicitly
(Equation~\ref{vofr}) and reexpress the density as follows:
\begin{equation}
n(R)=\left(\frac{\dot{M}}{v_\infty}\right)(4\pi\mu)^{-1}
R^{-2}\left[\vovf+\left(1-\vovf\right)\left(1-\frac{R_{\rm in}}{R}\right)^\beta
\right]^{-1}.
\end{equation}
According to this parameterization, the parameter \movf ~plays the
role of $\eta$.  The shape of the DEM distribution is then then a
function of the parameters \lvom, $\beta$, $v_0/v_\infty$, and, for
the disk wind, $R_{\rm in}$.  For a given choice of those parameters,
the magnitude of the DEM distribution is proportional to
$(\dot{M}/v_\infty)^2$ or, equivalently, $L^2$.

\subsection{Calculation of DEM Distributions and Global Spectra}
\label{calcdem}

The wind distributions we use are symmetric to rotation around the
line containing the neutron star and the center of the companion star.
This allowed us to conserve computational resources by using rings
around this axis as the spatial cells to calculate the DEM distributions.
The exclusion of the region occulted by the companion star from the
apparent DEM breaks this rotational symmetry but we were still able
to conserve computation, using rings as the spatial cells, by
computing the fraction of each ring not occulted by the companion star
and multiplying the emission measure for each ring by that
factor.  We used the binary system parameters from
Table~\ref{syspars}, and fixed $v_0/v_\infty$ at 0.015 which
corresponds approximately to the photospheric thermal velocity of an
O-type star for $v_0$ if $v_\infty$ is of order 1000\,km\,s$^{-1}$.
For the spectra which were accumulated during an orbital phase
interval longer than 0.02, we averaged the apparent DEM over phases
separated by no more than that.  Our parameter grid contains the
values $\beta=(0.0, 0.1,0.2,..., 1.5)$,
$\log(Lv_\infty/\dot{M})=(-0.5, -0.33, -0.17, 0, 0.17,..., 2.5)$ in
units of $10^{37}$\,\ergs ~for $L$, 1000\,km\,s$^{-1}$ for $v_\infty$
and $10^{-6}\,M_\sun$\,yr$^{-1}$ for $\dot{M}$.  For the disk wind,
the grid also contained $R_{\rm in}=(3.16, 5.62, 10,
17.8,31.6)\,R_\sun$.  The resultant spectra were stored in a FITS
format XSPEC ATABLE file \citep{arn95}.

In order to demonstrate the appearance and behavior of the DEM
distributions, we plot DEM distributions and contour maps of $\log\xi$
using the best fit parameters found in the following section
(\sect~\ref{fitting}, Table~\ref{fitdata}).  In Figure~\ref{spider} we
show a map of $\log\xi$ for the companion star wind and in
Figure~\ref{star_dems} we plot the apparent DEM distributions for
orbital phases representative of the observed phases.  In
Figures~\ref{spider_disk}~\&~\ref{disk_dems} we show the same plots
for the disk wind model.  However, as mentioned in
\sect~\ref{dem_mods}, we expect the Roche lobe radius to be an upper
limit on $R_{\rm in}$ and so we use $R_{\rm in}=3.4R_\sun$, the Roche
lobe radius as determined from the parameters of Table~\ref{syspars}
and the formula of \citet{egg83}, for these plots instead of a best
fit value of $R_{\rm in}$.  In Figure~\ref{disk_dems}, it can be seen
that the apparent emission measure is dramatically reduced during
eclipse for a disk wind inner radius significantly smaller than the
companion star radius.

\subsection{Spectral Fitting}
\label{fitting}

For both the disk and the stellar wind models, we simultaneously fit
the spectra at all four phases, using for $f_{\rm plasma}$ the
recombination spectrum for the observable wind at each phase as
discussed above.  We allowed the parameters of all of the spectral
components except for $f_{\rm plasma}$ to vary independently for the
four observed phases.  We show the best fit values and associated
errors in Table~\ref{fitdata}.  We show the spectral fits for the
stellar wind in Figure~\ref{four_fit}. The best fit model spectrum for
the eclipse phase is shown at high resolution in
Figure~\ref{ecl_unconvolved}.  Since our focus is on the line
spectrum, we let the continuum parameters vary to fit the continuum
shape without constraining them to physically meaningful bounds.  Some
strange results were obtained, such as the very large columns,
normalizations, and photon indexes for the second power law in the
egress phase (for both the stellar and disk wind models).  Though the
normalizations are very large, the fluxes of this component are
comparable to that in the other phases.  We note that our best fit
value of the neutron star luminosity from the stellar wind model is
very near the value obtained by direct measurement of the unocculted
continuum.  We emphasize that this luminosity value is obtained only
from our recombination spectra for wind models, and is determined
almost entirely by the observed line fluxes, not from any direct
measurement of the broad band flux.

Both wind models provide acceptable values of $\chi^2$, though for the
disk wind the fits favor values of the inner radius which are at
least comparable to the radius of the companion star.  In
Figure~\ref{diskchi}, we plot the value of $\chi^2$ for the fits as a
function of the assumed inner radius.  As mentioned in
\sect~\ref{calcdem}, values of $R_{\rm in}$ much smaller than the
companion star radius lead to dramatic variations in the DEM
distribution, and therefore the emission line flux over eclipse.  The
observed emission line spectrum, however, varies only moderately
across the eclipse.  As the preferred radii are very large compared to
the neutron star Roche-lobe radius, we consider the possibility that
the wind could be dominated by matter arising from the accretion disk
unlikely.

We compare the lines fluxes predicted by our best fit stellar wind
model with line fluxes measured by \citet{ebi96} in
Figure~\ref{fig:line_fluxes}.  According to our model, the lines from
elements with lower values of $Z$ vary somewhat more as a function of
orbital phase.  This is because the lines from low-$Z$ ions are
produced more efficiently in the region near the surface of the
companion star, where the ionization parameter is lower, and which is
occulted more during eclipse.  Unfortunately, the quality of the
current data is not sufficient to test this prediction.  \citet{ebi96}
noted that the H-like and He-like lines of iron were undetectable
during the first phase (``pre-eclipse'') from which they inferred that
the overall ionization parameter was lower at that phase.  Our
analysis presumes that we observe the same symmetric wind, only
occulted differently, at every phase.  Therefore, in our model, the
line fluxes must be the same at symmetric phases.  However, our
predicted line fluxes are greater only by a factor of a few from their
$1\sigma$ limits and so our assumption that the wind density
distribution and luminosity are constant is not excluded.

\section{Optical Depth of the Stellar Wind}
\label{optdep}

Until now, we have assumed that the stellar wind is optically thin to
X-rays from the neutron star.  For the densities and luminosities of
our best fit model, the state of X-ray photoionized gas is affected
little by optical depth inside an ionized Str\"omgren-type zone, where
the charge state distribution of helium is dominated by He$^{++}$ but
outside this zone, where the charge state distribution of helium
becomes dominated by He$^+$, the effects of optical depth quickly
become important \citep[e.g., ][]{kal82a}.  The size of a Str\"omgren
zone is usually derived for spherically symmetric nebulae \citep[e.g.,
][]{ost89} with the condition that the rate at which the central
source emits photons capable of ionizing helium is equal to the total
rate of recombinations which do not produce ionizing photons (i.e.,
recombinations to excited states).  However, the Str\"omgren radius
may also be estimated in a nebula which is not spherically symmetric
for any direction from the ionizing source using the condition that
the rate of helium ionizing photons emitted per solid angle is equal
to the total rate of recombinations per solid angle:
\begin{equation}
\frac{1}{4\pi}\int_{\nu_0}^\infty\frac{L_\nu}{h\nu}\dif\nu=
\int_0^{R_{\rm S}}\alpha_{\rm B}n_{\rm e} n_{\rm He^{++}}n_{\rm e} r^2 \dif r 
\label{strom}
\end{equation}
where $\alpha_{\rm B}$ is the recombination rate to excited states and
$\nu_0$ is ionization threshold of He$^{++}$ ($h\nu_0=54.4$\,eV).  If
we assume that the spectrum of Cen~X-3 is a power-law with photon
index 1, cut off at energies above $E_{\rm cut}=15$\,keV
(approximately the \citealt{san98} spectrum) then the integral on the
left hand side is
\begin{equation}
L E_{\rm cut}^{-1}\ln(E_{\rm cut}/h\nu_0)=\frac{L}{2.67\,{\rm
keV}}=2.3\ee{46}\,{\rm s}^{-1}  
\end{equation}
for $L=10^{38}$\,\ergs.
The computation of the integral on the right-hand side is complicated
by the fact that the recombination coefficient, $\alpha_{\rm B}$, is
a function of temperature.  However, the temperature dependence of
$\alpha_{\rm B}$ is only $\sim T^{-1/2}$ and so we can get a good
estimate by assuming that the nebula is isothermal.  In the
calculations of \cite{kal82a}, the temperature just inside of the
He$^{++}$/He$^{+}$ boundary is approximately $10^5$\,K and so we use
for $\alpha_{\rm B}$ the constant value $2\ee{-13}$\,cm$^3$\,s$^{-1}$
\citep[][Table 2.8]{ost89}.  In the ionized zone, nearly all of the
helium is in the 
form He$^{++}$ and so $n_{\rm He^{++}}=8.3\ee{-2}n_{\rm e}$
almost exactly.  The condition that the nebula is
optically thin becomes:
\begin{equation}
\int_0^R n_e^2 r^2 dr<1.1\ee{59}\,{\rm cm}^{-3}.
\end{equation}
We note that the quantity on the left hand side is the emission
measure per solid angle.  For our wind, from the neutron star to the
face of the companion along the line of centers, this integral is
$2.72\ee{57}$\,cm$^{-3}$ and in the direction away from the companion
to infinity, it is $2.98\ee{56}$\,cm$^{-3}$.  Therefore, it is clear
that the assumption that our model wind is optically thin, which we
have used to calculate emission spectra, is self-consistent.

\section{Discussion}
\label{discussion}

We note that our results for the global wind models can be generalized
beyond the homogeneous, spherically symmetric distributions we have
used.  If the wind is clumped on a scale which is small compared to
the system such that the ionization parameter inside the clumps and
the total emission measure is the same as in a corresponding
homogeneous model, then the observed spectrum will be unchanged.  To
construct such a model where the clumps have a filling factor $f$, the
density must be increased by a factor $f^{-1/2}$ to keep $\int n_{\rm e}^2
\dif V$ unchanged and so to keep $\xi$ unchanged, $L$ must also be
increased by a factor $f^{-1/2}$.  Because our best fit luminosity is
already at the high end of the range of luminosities observed for
Cen~X-3, the wind cannot be clumped with a filling factor
significantly smaller than unity.  A similar, though no longer exact,
extension can be made to winds that, instead of being confined in
small clumps, are confined in solid angle.  An example of this
scenario is a wind which exists on the surface a cone.  Such a
geometry is believed to exist in broad absorption line quasars and may be
due to preferential launching of the wind along rays from the
central object which graze the surface of the accretion 
disk \citep[e.g., ][]{mur95,pro00}.  For such a
geometry, the factor $\Omega/4\pi$, where $\Omega$ is the solid angle
to which the wind is confined, plays the same role as $f$ above.  For
the same reasons then, we can exclude these types of winds.

Both the stellar wind and the disk wind models give statistically
satisfactory fits to the data.  However, the best fit values of the
inner radius in the disk model are very large compared to the size of
the accretion disk.  If the extended reprocessing material originates
primarily from an accretion disk wind, a geometry in which the wind is
prevented from producing recombination radiation inside a radius which
is approximately equal to companion star radius (which is $\sim$4
times the maximum size of the accretion disk) is required.  We believe
that it would require fine-tuned conditions to produce such a
geometry, and we consider the possibility that a disk wind dominates
the extended circumstellar material unlikely.

As we have noted, in our analysis, the mass loss rate and terminal
velocity can be obtained only in the combination $\dot{M}/v_\infty$.
If  Doppler shifts of lines due to motion of the wind could be
measured, these two parameters could be determined independently.  The
errors in the line energies quoted by \citet{ebi96} correspond to
velocity upper limits of $\sim$2000\,\kms ~for the iron lines and
$\sim$6000\,\kms ~for the lower energy lines.  Therefore, only
an upper limit on the mass loss rate of the highly ionized wind
of $\sim3\ee{-6}\,M_\sun$\,yr$^{-1}$ can be obtained.  

The fact that the value of the luminosity that we obtain from our
spectral line fits is very close to the value determined from direct
measurements of the continuum gives us confidence that the parameters
of our model correspond to unique physical values that accurately
describe the stellar wind.  For the stellar wind, the best fit values
of $\dot{M}/v_\infty$ and $\beta$ are similar to those found in
isolated stars.  For example, for the 06.5\,Iaf star \objectname{HD
163758} and the 06.5\,III(f) star \objectname{HD 190864}, the values
of $\dot{M}$ are $6\ee{-6}$ and $1.5\ee{-6}$ $M_\sun$\,yr$^{-1}$ and
the values of $v_\infty$ are 2200 and 2500\,km\,s$^{-1}$ \citep{lam99}
which corresponds to values of $\dot{M}/v_\infty$ of
2.7$\ee{-9}\,M_\sun$\,yr$^{-1}$\,(\kms)$^{-1}$ and
0.6$\ee{-9}\,M_\sun$\,yr$^{-1}$\,(\kms)$^{-1}$ as compared with
1.56$\pm0.12\ee{-9}\,M_\sun$\,yr$^{-1}$\,(\kms)$^{-1}$
(Table~\ref{fitdata}) for our fits to the Cen~X-3 spectrum.  The
values of $\beta$ for these two isolated O stars are 0.7 and 0.8
\citep{lam99}, which are also near our derived value of
0.57\ud{0.06}{0.07} (Table~\ref{fitdata}).  Therefore, in our model,
the best fit wind parameters are roughly consistent with those of the
normal radiatively driven winds in isolated massive stars.
Considering that the wind is highly ionized and the radiation driving
mechanism which governs the structure of the wind in isolated massive
stars cannot function, this result is very surprising.

To within an order of magnitude or so, the total emission measure in a
smooth wind is given by $E \sim (\dot{M}/v_{\infty})^2/(m_{\rm p}^2
R_*)$.  Therefore, to the extent that the companion stars in HMXBs are
roughly the same size, lose mass at roughly the same rate, and produce
smooth winds with roughly the same terminal velocities, the total
emission measures should be approximately equal.  Numerically, the
above estimate gives, for typical parameters, $E \sim
10^{59}~\rm{cm^{-3}}$. (Note that this estimate applies equally well
to isolated early-type stars.) Clumping in the wind, however, will
tend to increase the emission measure. On the other hand, clumping
also tends to decrease the local value of $\xi$, in effect, removing
gas from the high-$\xi$ end of the DEM distribution and adding gas to
the lower-$\xi$ end. Thus clumping reduces the wind's X-ray
recombination line emission, and increases its fluorescence emission.

We have argued that the Cen~X-3 wind is smooth. \citet{sak99} have
shown that the wind in Vela X-1 is highly clumped. In
Figure~\ref{vc_logdem}, we compare the DEM distribution derived here
for Cen~X-3 with that derived, also using X-ray recombination, by
\citet{sak99} for Vela~X-1.  The value of $\dot{M}/v_{\infty}$ we
have derived here for the wind of Cen~X-3 is comparable to the value
for Vela~X-1 derived by a number of methods \citep{dup80,sat86,sak99}.
As one caveat, we point out that for Vela X-1 the contribution from
fluorescing material, which would cause an upturn in its DEM curve
below $\log \xi \approx 1.5$, is not included in the
plot.\footnote{Mapping the DEM distribution for low-$\xi$ material
based upon X-ray spectra requires much higher spectral resolution, so
that fluorescent line complexes can be resolved into their respective
charge states.}  Above $\log \xi=2.0$, however, the differences
are real, and are quite striking. The relatively small DEM magnitudes
in Vela X-1 are a consequence of the fact that most of the mass in the
wind is ``locked up'' in clumps of high density, hence, low $\xi$.
For example, near $\log \xi=3.0$, where lines from He-like and H-like
iron are produced, the DEM magnitudes vary by $\sim100$, which means
that these lines are $\sim100$ times more luminous in Cen~X-3 than in
Vela X-1. This does not translate into excessively large line
equivalent widths in Cen~X-3, however, since its X-ray luminosity is
also roughly 100 times higher.
 
With this comparison arises the question as to what conditions are
required in order to produce and/or destroy clumps in X-ray irradiated
winds, a subject beyond the scope of this paper, but possibly one that
bears on the nature of the wind driving force.  The absence of
substantial clumping in Cen~X-3, as inferred from the modest phase
variations of the iron K$\alpha$ equivalent width, is demonstrated by,
and is consistent with, the large DEM magnitudes for $\log \xi >
2$. It has been suggested \citep{sak99} that the existence of a
clumped wind component in Vela X-1 allows normal radiative driving by
the UV field of the companion. Clearly, that is disallowed in the case
of Cen~X-3.  We therefore conclude that the wind is most likely driven
by X-ray heating of the illuminated surface of the companion star as
proposed by \citet{day93a}.

We look forward to {\it Chandra} observations of this system.  High
resolution spectroscopic data will allow us to measure independently
the lines of the helium-like 2$\rightarrow$1 triplets and therefore
determine the ionization mechanism directly \citep{lie99}.  We will
also be able to measure much smaller velocities (of order
100\,km\,s$^{-1}$) and therefore be able to set much tighter
constraints on the mass-loss rate of the companion and the dynamics of
the wind.

\appendix

\section{Differential Emission Measure}
\label{app:dem}

We have defined the emission measure as $\int n_{\rm e}^2 \dif V$.
We now define the ionization parameter limited emission measure as:
\begin{equation}
\label{deflem}
E(\xi)\equiv\int_{V(\xi^\prime\leq\xi)} n_{\rm e}^2 \dif V,
\end{equation}
i.e., the emission measure in the volume in which the ionization
parameter is less than some value.  The linear differential emission
measure is then
\begin{equation}
\label{defdem}
\frac{\dif E(\xi)}{\dif \xi}=\lim_{\delta\xi\to 0} \frac{E(\xi +\delta \xi)-E(\xi)}{\delta\xi}
=\frac{1}{\delta \xi}\int_{V(\xi\leq\xi^\prime\leq\xi+\delta\xi)}
n_{\rm e}^2 \dif V.
\end{equation}
This infinitesimal volume is the space between
two surfaces, each of which is defined by a single value of the
ionization parameter.
The infinitesimal distance between the surfaces is
$\delta\xi/|\nabla\xi|$.  Therefore,
\begin{equation}
\frac{\dif E(\xi)}{\dif \xi}=\int_{S(\xi)}\frac{n_{\rm e}^2}{|\nabla \xi|}\,\dif S.
\end{equation}
where $S(\xi)$ and $dS$ specify the integral over the surface
specified by the ionization parameter $\xi$.  We define the
differential emission measure
\begin{equation}
\label{dems}  
D(\xi)\equiv\frac{\dif E(\xi)}{\dif \log\xi}=
\int_{S(\xi)}\frac{n_{\rm e}^2}{|\nabla \log\xi|}\,\dif S.
\end{equation}

Suppose we have a model for the density distribution such that the
density scales linearly with the parameter $\eta$.  Consider a change
of parameters $\eta\to \eta^\prime$ and $L\to L^\prime$.  The surface
defined by $\xi$ for the unprimed parameters is the same surface
defined by
\begin{equation}
\xi^\prime=\frac{L^\prime \eta}{L\eta^\prime}\xi.
\end{equation}
Therefore, $D^\prime(\xi^\prime)$ can be related to
$D(\xi)$ by transforming the integrand 
in Equation~\ref{dems}.  Under the parameter change, $\log\xi$ differs
from $\log\xi^\prime$ by an additive constant and so
$|\nabla\log\xi|=|\nabla\log\xi^\prime|$.  However, $n_{\rm e}^2$ is
modified by the factor $(\eta^\prime/\eta)^2$.  Therefore,
\begin{equation}
D^\prime(\xi^\prime)=\left(\frac{\eta^\prime}{\eta}\right)^2D(\xi).
\end{equation}
This is Equation~\ref{dem_scal}.

\acknowledgements

  We thank Chris Mauche and Daniel Proga for helpful discussions and
careful reading of the manuscript.  We also thank John Blondin for
helpful discussions.  This research has made use of data obtained
through the High Energy Astrophysics Science Archive Research Center
Online Service, provided by the NASA/Goddard Space Flight Center.
This research has made use of NASA's Astrophysics Data System Abstract
Service.  D.\ A.\ L.\ was supported in part by NASA Long Term Space
Astrophysics grant S-92654-F.  Work at LLNL was performed under the
auspices of the U.\ S.\ Department of Energy by University of
California Lawrence Livermore National Laboratory under Contract
W-7405-Eng-48.  M.\ S.\ was supported under NASA grant NAG 5-7737.

\bibliographystyle{apj} 
\bibliography{ms}

\begin{thebibliography}{67}
\expandafter\ifx\csname natexlab\endcsname\relax\def\natexlab#1{#1}\fi

\bibitem[{{Alme} \& {Wilson}(1974)}]{alm74}
{Alme}, M.~L. \& {Wilson}, J.~R. 1974, \apj, 194, 147

\bibitem[{{Anders} \& {Grevesse}(1989)}]{and89}
{Anders}, E. \& {Grevesse}, N. 1989, \gca, 53, 197

\bibitem[{Arnaud(1995)}]{arn95}
Arnaud, K.~A. 1995, The File Format for XSPEC Table Models, Office of Guest
  Investigator Programs Memo OGIP/92-009, NASA Goddard Space Flight Center
  Laboratory for High Energy Astrophysics

\bibitem[{Arnaud(1996)}]{arn96}
Arnaud, K.~A. 1996, in ASP Conf. Series, Vol. 101, Astronomical Data Analysis
  Software and Systems V, ed. G.~Jacoby \& J.~Barnes, 17

\bibitem[{{Arons}(1973)}]{aro73}
{Arons}, J. 1973, \apj, 184, 539

\bibitem[{{Ash} {et~al.}(1999){Ash}, {Reynolds}, {Roche}, {Norton}, {Still}, \&
  {Morales-Rueda}}]{ash99}
{Ash}, T. D.~C., {Reynolds}, A.~P., {Roche}, P., {Norton}, A.~J., {Still},
  M.~D., \& {Morales-Rueda}, L. 1999, \mnras, 307, 357

\bibitem[{{Audley}(1998)}]{aud98}
{Audley}, M.~D. 1998, PhD thesis, Univ. of Maryland

\bibitem[{{Basko} \& {Sunyaev}(1973)}]{bas73}
{Basko}, M.~M. \& {Sunyaev}, R.~A. 1973, \apss, 23, 117

\bibitem[{{Becker} {et~al.}(1978){Becker}, {Pravdo}, {Rothschild}, {Boldt},
  {Holt}, {Serlemitsos}, \& {Swank}}]{bec78}
{Becker}, R.~H., {Pravdo}, S.~H., {Rothschild}, R.~E., {Boldt}, E.~A., {Holt},
  S.~S., {Serlemitsos}, P.~J., \& {Swank}, J.~H. 1978, \apj, 221, 912

\bibitem[{{Blandford} \& {Payne}(1982)}]{bla82}
{Blandford}, R.~D. \& {Payne}, D.~G. 1982, \mnras, 199, 883

\bibitem[{{Blondin} \& {Woo}(1995)}]{blo95}
{Blondin}, J.~M. \& {Woo}, J.~W. 1995, \apj, 445, 889

\bibitem[{{Burderi} {et~al.}(2000){Burderi}, {Di Salvo}, {Robba}, {La Barbera},
  \& {Guainazzi}}]{bur00}
{Burderi}, L., {Di Salvo}, T., {Robba}, N.~R., {La Barbera}, A., \&
  {Guainazzi}, M. 2000, \apj, 530, 429

\bibitem[{{Castor} {et~al.}(1975){Castor}, {Abbott}, \& {Klein}}]{cas75}
{Castor}, J.~I., {Abbott}, D.~C., \& {Klein}, R.~I. 1975, \apj, 195, 157

\bibitem[{{Clark} {et~al.}(1988){Clark}, {Minato}, \& {Mi}}]{cla88}
{Clark}, G.~W., {Minato}, J.~R., \& {Mi}, G. 1988, \apj, 324, 974

\bibitem[{{Conti}(1978)}]{con78}
{Conti}, P.~S. 1978, \aap, 63, 225

\bibitem[{{Day} {et~al.}(1993){Day}, {Nagase}, {Asai}, \& {Takeshima}}]{day93b}
{Day}, C. S.~R., {Nagase}, F., {Asai}, K., \& {Takeshima}, T. 1993, \apj, 408,
  656

\bibitem[{{Day} \& {Stevens}(1993)}]{day93a}
{Day}, C. S.~R. \& {Stevens}, I.~R. 1993, \apj, 403, 322

\bibitem[{{Dupree} {et~al.}(1980){Dupree} {et~al.}}]{dup80}
{Dupree}, A.~K., {et~al.} 1980, \apj, 238, 969

\bibitem[{{Ebisawa} {et~al.}(1996){Ebisawa}, {Day}, {Kallman}, {Nagase},
  {Kotani}, {Kawashima}, {Kitamoto}, \& {Woo}}]{ebi96}
{Ebisawa}, K., {Day}, C. S.~R., {Kallman}, T.~R., {Nagase}, F., {Kotani}, T.,
  {Kawashima}, K., {Kitamoto}, S., \& {Woo}, J.~W. 1996, \pasj, 48, 425

\bibitem[{{Eggleton}(1983)}]{egg83}
{Eggleton}, P.~P. 1983, \apj, 268, 368

\bibitem[{{Friend} \& {Castor}(1982)}]{fri82}
{Friend}, D.~B. \& {Castor}, J.~I. 1982, \apj, 261, 293

\bibitem[{{Gendreau}(1995)}]{gen95}
{Gendreau}, K.~C. 1995, PhD thesis, Massachusetts Institute of Technology

\bibitem[{{Haberl} {et~al.}(1989){Haberl}, {White}, \& {Kallman}}]{hab89}
{Haberl}, F., {White}, N.~E., \& {Kallman}, T.~R. 1989, \apj, 343, 409

\bibitem[{{Hammerschlag-Hensberge} {et~al.}(1984){Hammerschlag-Hensberge},
  {Kallman}, \& {Howarth}}]{ham84}
{Hammerschlag-Hensberge}, G., {Kallman}, T.~R., \& {Howarth}, I.~D. 1984, \apj,
  283, 249

\bibitem[{{Hatchett} \& {McCray}(1977)}]{hat77}
{Hatchett}, S. \& {McCray}, R. 1977, \apj, 211, 552

\bibitem[{{Ho} \& {Arons}(1987)}]{ho87}
{Ho}, C. \& {Arons}, J. 1987, \apj, 316, 283

\bibitem[{{Hutchings} {et~al.}(1979){Hutchings}, {Cowley}, {Crampton}, {Van
  Paradus}, \& {White}}]{hut79}
{Hutchings}, J.~B., {Cowley}, A.~P., {Crampton}, D., {Van Paradus}, J., \&
  {White}, N.~E. 1979, \apj, 229, 1079

\bibitem[{Kallman \& {Krolik}(1999)}]{kal99}
Kallman, T.~R. \& {Krolik}, J.~H. 1999, XSTAR v1.43, HEASARC (NASA/GSFC),
  Greenbelt, MD

\bibitem[{{Kallman} \& {McCray}(1982)}]{kal82a}
{Kallman}, T.~R. \& {McCray}, R. 1982, \apjs, 50, 263

\bibitem[{{Kallman} \& {White}(1982)}]{kal82b}
{Kallman}, T.~R. \& {White}, N.~E. 1982, \apjl, 261, L35

\bibitem[{{Kawashima} \& {Kitamoto}(1996)}]{kaw96}
{Kawashima}, K. \& {Kitamoto}, S. 1996, \pasj, 48, L113

\bibitem[{{Kitamoto} {et~al.}(1994){Kitamoto}, {Kawashima}, {Negoro},
  {Miyamoto}, {White}, \& {Nagase}}]{kit94}
{Kitamoto}, S., {Kawashima}, K., {Negoro}, H., {Miyamoto}, S., {White}, N.~E.,
  \& {Nagase}, F. 1994, \pasj, 46, L105

\bibitem[{Klapisch {et~al.}(1977)Klapisch, Schwab, Fraenkel, \& Oreg}]{kla77}
Klapisch, M., Schwab, J.~L., Fraenkel, J.~S., \& Oreg, J. 1977, J. Opt. Soc.
  Am., 61, 148

\bibitem[{{Kudritzki} {et~al.}(1989){Kudritzki}, {Pauldrach}, {Puls}, \&
  {Abbott}}]{kud89}
{Kudritzki}, R.~P., {Pauldrach}, A., {Puls}, J., \& {Abbott}, D.~C. 1989, \aap,
  219, 205

\bibitem[{{Lamers} {et~al.}(1999){Lamers}, {Haser}, {De Koter}, \&
  {Leitherer}}]{lam99}
{Lamers}, H. J. G. L.~M., {Haser}, S., {De Koter}, A., \& {Leitherer}, C. 1999,
  \apj, 516, 872

\bibitem[{Liedahl(1999)}]{lie99}
Liedahl, D.~A. 1999, in Lecture Notes in Physics, Vol. 520, X-Ray Spectroscopy
  in Astrophysics, Lectures held at the Tenth Summer School of the European
  Astrophysics Doctoral Network in Amsterdam, the Netherlands. September 22 -
  October 3, 1997, ed. J.~{van Paradijs} \& J.~A.~M. Bleeker (New York:
  Springer-Verlag), 189

\bibitem[{Liedahl \& Paerels(1996)}]{lie96}
Liedahl, D.~A. \& Paerels, F. 1996, \apj, 468, L33

\bibitem[{{Lucy} \& {Solomon}(1970)}]{luc70}
{Lucy}, L.~B. \& {Solomon}, P.~M. 1970, \apj, 159, 879

\bibitem[{{MacGregor} \& {Vitello}(1982)}]{mgr82}
{MacGregor}, K.~B. \& {Vitello}, P. A.~J. 1982, \apj, 259, 267

\bibitem[{{Masai}(1984)}]{mas84}
{Masai}, K. 1984, \apss, 106, 391

\bibitem[{{McCray} \& {Hatchett}(1975)}]{mcr75}
{McCray}, R. \& {Hatchett}, S. 1975, \apj, 199, 196

\bibitem[{Mewe {et~al.}(1996)Mewe, Kaastra, \& Liedahl}]{mew95}
Mewe, R., Kaastra, J.~S., \& Liedahl, D.~A. 1996, Legacy, 6, 16

\bibitem[{{Morrison} \& {McCammon}(1983)}]{mor83}
{Morrison}, R. \& {McCammon}, D. 1983, \apj, 270, 119

\bibitem[{{Morton}(1967)}]{mor67}
{Morton}, C.~D. 1967, \apj, 150, 535

\bibitem[{{Murray} {et~al.}(1995){Murray}, {Chiang}, {Grossman}, \&
  {Voit}}]{mur95}
{Murray}, N., {Chiang}, J., {Grossman}, S.~A., \& {Voit}, G.~M. 1995, \apj,
  451, 498

\bibitem[{{Nagase} {et~al.}(1992){Nagase}, {Corbet}, {Day}, {Inoue},
  {Takeshima}, {Yoshida}, \& {Mihara}}]{nag92}
{Nagase}, F., {Corbet}, R. H.~D., {Day}, C. S.~R., {Inoue}, H., {Takeshima},
  T., {Yoshida}, K., \& {Mihara}, T. 1992, \apj, 396, 147

\bibitem[{{Nagase} {et~al.}(1986){Nagase}, {Hayakawa}, {Sato}, {Masai}, \&
  {Inoue}}]{nag86}
{Nagase}, F., {Hayakawa}, S., {Sato}, N., {Masai}, K., \& {Inoue}, H. 1986,
  \pasj, 38, 547

\bibitem[{{Nagase} {et~al.}(1994){Nagase}, {Zylstra}, {Sonobe}, {Kotani},
  {Inoue}, \& {Woo}}]{nag94}
{Nagase}, F., {Zylstra}, G., {Sonobe}, T., {Kotani}, T., {Inoue}, H., \& {Woo},
  J. 1994, \apjl, 436, L1

\bibitem[{{Osterbrock}(1989)}]{ost89}
{Osterbrock}, D.~E. 1989, "Astrophysics of gaseous nebulae and active galactic
  nuclei" (Mill Valley, CA: University Science Books)

\bibitem[{{Owen} \& {Blondin}(1997)}]{owe97}
{Owen}, M.~P. \& {Blondin}, J.~M. 1997, in American Astronomical Society
  Meeting, Vol. 190, 4505

\bibitem[{{Proga}(2000)}]{pro00}
{Proga}, D. 2000, \apj, accepted

\bibitem[{{Proga} {et~al.}(1999){Proga}, {Stone}, \& {Drew}}]{pro99}
{Proga}, D., {Stone}, J.~M., \& {Drew}, J.~E. 1999, \mnras, 310, 476

\bibitem[{{Sako} {et~al.}(1999){Sako}, {Liedahl}, {Kahn}, \& {Paerels}}]{sak99}
{Sako}, M., {Liedahl}, D.~A., {Kahn}, S.~M., \& {Paerels}, F. 1999, \apj, 525,
  921

\bibitem[{Saloman {et~al.}(1988)Saloman, Hubble, \& Scofield}]{sal88}
Saloman, E.~B., Hubble, J.~H., \& Scofield, J.~H. 1988, At. Data Nucl. Data
  Tables, 38

\bibitem[{{Santangelo} {et~al.}(1998){Santangelo}, {Del Sordo}, {Segreto}, {Dal
  Fiume}, {Orlandini}, \& {Piraino}}]{san98}
{Santangelo}, A., {Del Sordo}, S., {Segreto}, A., {Dal Fiume}, D., {Orlandini},
  M., \& {Piraino}, S. 1998, \aap, 340, L55

\bibitem[{{Sato} {et~al.}(1986){Sato}, {Hayakawa}, {Nagase}, {Masai}, {Dotani},
  {Inoue}, {Makino}, {Makishima}, \& {Ohashi}}]{sat86}
{Sato}, N., {Hayakawa}, S., {Nagase}, F., {Masai}, K., {Dotani}, T., {Inoue},
  H., {Makino}, F., {Makishima}, K., \& {Ohashi}, T. 1986, \pasj, 38, 731

\bibitem[{Schreier {et~al.}(1972)Schreier, Levinson, Gursky, Kellogg,
  Tananbaum, \& Giacconni}]{sch72}
Schreier, E., Levinson, R., Gursky, H., Kellogg, E., Tananbaum, H., \&
  Giacconni, R. 1972, \apj, 172, L79

\bibitem[{{Serlemitsos} {et~al.}(1995){Serlemitsos}, {Jalota}, {Soong},
  {Kunieda}, {Tawara}, {Tsusaka}, {Suzuki}, {Sakima}, {Yamazaki}, {Yoshioka},
  {Furuzawa}, {Yamashita}, {Awaki}, {Itoh}, {Ogasaka}, {Honda}, \&
  {Uchibori}}]{ser95}
{Serlemitsos}, P.~J., {Jalota}, L., {Soong}, Y., {Kunieda}, H., {Tawara}, Y.,
  {Tsusaka}, Y., {Suzuki}, H., {Sakima}, Y., {Yamazaki}, T., {Yoshioka}, H.,
  {Furuzawa}, A., {Yamashita}, K., {Awaki}, H., {Itoh}, M., {Ogasaka}, Y.,
  {Honda}, H., \& {Uchibori}, Y. 1995, \pasj, 47, 105

\bibitem[{{Stevens}(1991)}]{ste91}
{Stevens}, I.~R. 1991, \apj, 379, 310

\bibitem[{Tarter {et~al.}(1969)Tarter, Tucker, \& Salpeter}]{tar69}
Tarter, C.~B., Tucker, W.~H., \& Salpeter, E.~E. 1969, \apj, 156, 943

\bibitem[{{The Ftools Group}(1998)}]{ft42}
{The Ftools Group}. 1998, FTOOLS v4.2, High Energy Astrophysics Science Archive
  Reserch Center, NASA Goddard Space Flight Center

\bibitem[{{van der Klis} {et~al.}(1982){van der Klis},
  {Hammerschlag-Hensberge}, {van Paradijs}, {Bonnet-Bidaud}, {Ilovaisky},
  {Mouchet}, {Chevalier}, {Glencross}, {Willis}, \& {Zuiderwijk}}]{vdk82}
{van der Klis}, M., {Hammerschlag-Hensberge}, G., {van Paradijs}, J.~A.,
  {Bonnet-Bidaud}, J.~M., {Ilovaisky}, S.~A., {Mouchet}, M., {Chevalier}, C.,
  {Glencross}, W.~M., {Willis}, A.~J., \& {Zuiderwijk}, E.~J. 1982, \aap, 106,
  339

\bibitem[{{Vrtilek} {et~al.}(1997){Vrtilek}, {Boroson}, {Cheng}, {McCray}, \&
  {Nagase}}]{vrt97}
{Vrtilek}, S.~D., {Boroson}, B., {Cheng}, F.~H., {McCray}, R., \& {Nagase}, F.
  1997, \apj, 490, 377+

\bibitem[{{Wojdowski} {et~al.}(2000){Wojdowski}, {Clark}, \& {Kallman}}]{woj00}
{Wojdowski}, P.~S., {Clark}, G.~W., \& {Kallman}, T.~R. 2000, \apj, in press

\bibitem[{{Woo} {et~al.}(1995){Woo}, {Clark}, {Blondin}, {Kallman}, \&
  {Nagase}}]{woo95}
{Woo}, J.~W., {Clark}, G.~W., {Blondin}, J.~M., {Kallman}, T.~R., \& {Nagase},
  F. 1995, \apj, 445, 896

\bibitem[{{Woo} {et~al.}(1994){Woo}, {Clark}, {Day}, {Nagase}, \&
  {Takeshima}}]{woo94}
{Woo}, J.~W., {Clark}, G.~W., {Day}, C. S.~R., {Nagase}, F., \& {Takeshima}, T.
  1994, \apjl, 436, L5

\bibitem[{{Woods} {et~al.}(1996){Woods}, {Klein}, {Castor}, {McKee}, \&
  {Bell}}]{woo96}
{Woods}, D.~T., {Klein}, R.~I., {Castor}, J.~I., {McKee}, C.~F., \& {Bell},
  J.~B. 1996, \apj, 461, 767

\end{thebibliography}

\begin{deluxetable}{lccc}
\tablewidth{0pt}
\tablecaption{Fit Parameters for Eclipse Spectrum}
\tablehead{
\colhead{ } & \multicolumn{3}{c}{Spectral Line Model} \\
 \cline{2-4} \\
\colhead{ } & \colhead{ } & \colhead{Variable} & \colhead{ }  \\
\colhead{ } & \colhead{ } & \colhead{Abundance} & \colhead{ } \\
\colhead{Parameter} & \colhead{MEKAL} & \colhead{MEKAL} & \colhead{Photo} }
\startdata
$N_{\rm H1}$ ($10^{22}$\,cm$^{-2}$) 
 & 0.96\ud{0.03}{0.02}	& 0.91$\pm$0.02	& 1.03$\pm$0.04	\\
$E$ (10$^{58}$\,cm$^{-3}$)($d$/10 kpc)$^{-2}$
 & 3.62\ud{0.06}{0.08}	& 0.17\ud{0.09}{0.04}	& 2.58\ud{0.22}{0.27} \\
$\log\xi$ & \nodata	& \nodata & 3.19\ud{0.05}{0.07} \\
$kT$ (keV)\tablenotemark{a}
 & 7.5\ud{0.4}{0.5} 	& 8.9$\pm$0.5	& 0.41$\pm$0.01 \\
$Z/Z_\sun$ & 1(fixed)	& 39\ud{112}{\phn16}	& 1(fixed) \\
$\alpha_1$    &	1.5(fixed) & 1.5(fixed) 	& 1.50$\pm$0.09	\\
$I_{\rm pl1}$ (10$^{-3}$ s$^{-1}$\,cm$^{-2}$\,keV$^{-1}$)
 & $<0.11$ & 5.6(fixed) & 6.7$\pm$0.6	\\
$n_{\rm H2}$  ($10^{22}$\,cm$^{-2}$) 
 & 71\ud{10}{11}	& 74$\pm$14 & 50\ud{11}{20} \\
$\epsilon_{\rm line}$ (keV)
 & 6.41\ud{0.018}{0.020} & 6.393\ud{0.015}{0.017} & 6.388\ud{0.020}{0.015} \\
$I_{\rm line}$(10$^{-4}$\,ph\,cm$^{-2}$ s$^{-1}$) 	
 & 3.8\ud{0.9}{0.8}	& 4.8\ud{1.4}{0.9}	& 3.3\ud{1.2}{0.6} \\
$\alpha_2$ & 1.5$\pm$0.4 & 1.6\ud{0.5}{0.6}	& 1.5\ud{0.7}{0.4}	\\
$I_{\rm pl2}$ (10$^{-3}$\,ph s$^{-1}$\,cm$^{-2}$)  
 & 39\ud{57}{25} & 11\ud{18}{\phn7}	& 17\ud{79}{10}	\\
$\chi^2$/d.o.f.  & 176/170	& 146/170	& 176/169 \\
probability\tablenotemark{b}	& 35\%	& 90\%	& 33\% \\
\enddata
\tablenotetext{a}{In the case of the photoionized model, $kT$ is an
explicit function of $\log\xi$.  The error on the temperature is
estimated using a local linear approximation to $T(\log\xi)$.}
\tablenotetext{b}{Null hypothesis probability}
\label{io_mod_fit}
\end{deluxetable}

\begin{deluxetable}{llc}
\tablewidth{0pt}
\tablecaption{Assumed System Parameters for Cen~X-3/V779~Cen}
\tablehead{
\colhead{Parameter} & \colhead{Value} & \colhead{Reference} }
\startdata
$R_\star$(companion radius) & 11.8 $R_\sun$  &  a \\
$a$(orbital separation)     & 19.2 $R_\sun$  &  a \\
$i$(inclination angle)      & 70.2\arcdeg    &  a \\
Distance		    & 10 kpc	     &  b \\
\enddata
\tablerefs{(a) \citealt{ash99} (b) \citealt{hut79}}
\label{syspars}
\end{deluxetable}

\begin{deluxetable}{lccccccccc}
\tablewidth{0pt}
\rotate
\tabletypesize{\scriptsize}
\tablecaption{Global Wind Model Fit Parameters}
\tablehead{
  &  \multicolumn{4}{c}{Stellar Wind} & & \multicolumn{4}{c}{Disk Wind} \\
\cline{2-5} \cline{7-10} \\
\colhead{Parameter} & \colhead{Pre-eclipse} & \colhead{Ingress} &
\colhead{Eclipse} & \colhead{Egress} &
\colhead{ } & \colhead{Pre-eclipse} & \colhead{Ingress} &
\colhead{Eclipse} & \colhead{Egress} } 
\startdata
$N_{\rm H1}$ ($10^{22}$\,cm$^{-2}$) & 0.95$\pm$0.08 &
0.97\ud{0.02}{0.04} & 1.00\ud{0.02}{0.04} &
0.94\ud{0.07}{0.08} &
& 0.86$\pm0.08$ & 0.94\ud{0.05}{0.04} & 1.03$\pm{0.04}$ &
0.87\ud{0.07}{0.08} \\
$\alpha_1$    & 0.59\ud{0.17}{0.15}   & 0.96\ud{0.09}{0.06} &
1.42\ud{0.11}{0.04}   & -0.18\ud{0.07}{0.04} &
& 0.56\ud{0.16}{0.14} & 0.96\ud{0.10}{0.06} & 1.41\ud{0.12}{0.05} &
-0.19\ud{0.03}{0.07} \\
$I_{\rm pl1}$ (10$^{-3}$\,ph\,cm$^{-2}$ s$^{-1}$\,keV$^{-1}$)	& 
17.4\ud{3.0}{2.4} & 7.3\ud{0.8}{0.5} & 5.8\ud{0.7}{0.5} &
3.1\ud{0.4}{0.3} &
& 17.2\ud{2.5}{2.3} & 7.3\ud{0.9}{0.2} & 5.5\ud{0.7}{0.2} &
3.09\ud{0.16}{0.27} \\ 
$N_{\rm H2}$ ($10^{22}$\,cm$^{-2}$) &
35\ud{8}{6} & 140\ud{27}{32} & 50\ud{\phn8}{14} & 183\ud{35}{45} &
& 38\ud{9}{7} & 154\ud{28}{36} & 46\ud{15}{13} & 197\ud{21}{15} \\
$\epsilon_{\rm line}$ (keV) 		&       
6.374\ud{0.023}{0.027} & 6.371\ud{0.016}{0.022} &
6.391\ud{0.017}{0.015} & 6.390$\pm$0.020 &
& 6.376\ud{0.024}{0.028} & 6.371\ud{0.017}{0.022} &
6.392\ud{0.016}{0.015} & 6.391\ud{0.020}{0.021}  \\
$I_{\rm line}$(10$^{-4}$\,ph\,cm$^{-2}$ s$^{-1}$) 	&
23$\pm$5 & 40\ud{25}{13} & 3.3\ud{0.7}{0.6} &
2.2\ud{1.8}{0.5}$\ee{2}$ &
&  24$\pm$5 & 49\ud{18}{15} & 3.2\ud{0.8}{0.6} & 27\ud{12}{\phn6} \\
$\alpha_2$    	& 1.5\ud{0.5}{0.3} & 3.3\ud{0.6}{0.9} &
1.5\ud{0.4}{0.5} & 7.5\ud{0.9}{0.4} &
& 1.6\ud{0.5}{0.4} & 3.6\ud{0.9}{1.0} & 1.5\ud{0.5}{0.3} &
8.0\ud{0.1}{1.2} \\
$I_{\rm pl2}$ (10$^{-3}$\,ph\,cm$^{-2}$s$^{-1}$\,keV$^{-1}$) & 
2.7\ud{4.3}{1.3}$\ee{2}$ & 6.1\ud{46}{\phn5.4}$\ee{3}$ & 20\ud{35}{14} &
8.3\ud{620}{\phn\phn0.9}$\ee{7}$ &
& 3.3\ud{5.8}{1.9}$\ee{2}$ & 1.4\ud{6.1}{1.3}$\ee{3}$ & 18\ud{39}{13} &
2.6\ud{190}{\phn\phn1.5}$\ee{8}$ \\
$R_{\rm in} (R_\sun)$ & \multicolumn{4}{c}{\nodata} &
                     & \multicolumn{4}{c}{ $>$15} \\
$\beta$	& \multicolumn{4}{c}{ 0.57\ud{0.06}{0.07}} &
	& \multicolumn{4}{c}{ 0.51\ud{0.17}{0.07}} \\
$Lv_\infty/\dot{M}$ \tablenotemark{a}
	& \multicolumn{4}{c}{6.8\ud{2.0}{1.8}}  &
	& \multicolumn{4}{c}{15\ud{22}{\phn6}} \\
$\dot{M}/v_\infty$[$d$/(10 kpc)]$^{-1}$ \tablenotemark{a}
	& \multicolumn{4}{c}{ 1.56$\pm$0.12} &
	& \multicolumn{4}{c}{ 1.3\ud{0.2}{0.6} } \\
$L$[$d$/(10 kpc)]$^{-1}$ \tablenotemark{a} 
	& \multicolumn{4}{c}{  10.7\ud{2.9}{2.4} }  &
	& \multicolumn{4}{c}{  20$\pm9$ }  \\
$\chi^2$ & \multicolumn{4}{c}{ 552/684} &
	 & \multicolumn{4}{c}{ 550/683\tablenotemark{b}} \\
\enddata

\tablenotetext{a}{These three expressions are combinations of only two
non-degenerate model parameters.  The units for these
values are such that $\dot{M}$ has units of $10^{-6}M_\sun$\,yr$^{-1}$,
$v_\infty$ units of 1000\,km\,s$^{-1}$ and $L$ units of $10^{37}$
\ergs.}

\tablenotetext{b}{for $R_{\rm in}$=30 $R_\sun$}
\label{fitdata}
\end{deluxetable}

\begin{deluxetable}{lccccc}
\tablewidth{0pt}
\tablecaption{Stellar Wind Model $n=2\to1$ Line
Fluxes\tablenotemark{a}} 
\tablehead{
    & Line        & \multicolumn{4}{c}{ Phase Range } \\
\cline{3-6} \\
Ion & Energy(keV)\tablenotemark{b} & -0.31 -- -0.29 & -0.23 -- -0.08 & 
-0.08 -- 0.13 & 0.14 -- 0.20 }
\startdata
\ion{Ne}{10} & 1.022 & 4.7 & 2.7 & 1.9 & 3.0 \\
\ion{Mg}{12} & 1.472 & 1.7 & 1.0 & 0.7 & 1.0 \\
\ion{Si}{14} & 2.006 & 2.0 & 1.3 & 1.0 & 1.4 \\
\ion{S}{16}  & 2.621 & 1.7 & 1.2 & 0.9 & 1.2 \\
\ion{Fe}{25} & 6.667 & 4.9 & 3.6 & 3.2 & 3.6 \\
\ion{Fe}{26} & 6.966 & 2.8 & 2.2 & 2.0 & 2.3 \\
\enddata
\tablenotetext{a}{In units of $10^{-4}$\,ph\,cm$^{-2}$\,s$^{-1}$ are summed over line complexes and energies are for
centroids of complexes.} 
\tablenotetext{b}{For He-like ions, the centroid energy of
the line complex.}
\label{tab:line_fluxes}
\end{deluxetable}

\clearpage

\figcaption[f1.ps]{The fit to the eclipse spectrum using the
MEKAL collisionally ionized plasma model.  The crosses represent the
data.  The solid line represents the total model spectrum and the
dashed and dotted lines represent the components of the model
spectrum.  The MEKAL model (lines and bremsstrahlung continuum)
accounts for nearly all of the observed flux below $\sim$5\,keV.
Power law 2 and the 6.4\,keV line make up the remainer of the flux at
high energy.  No room for a Compton scattered continuum remains below
5\,keV. \label{ecl_mekal}}

\figcaption[f2.ps]{The eclipse spectrum fit with the MEKAL
model with variable abundance.  Here, the power law is larger than the
MEKAL component at low energies though the emission features are
approximately the same as in solar abundance fit. \label{ecl_mekalv}}

\figcaption[f3.ps]{The fit using the single-zone recombination
model.  The recombination line and RRC spectrum (second largest
component in 1--2 keV region) is plotted as a component
separate from the 0.4\,keV bremsstrahlung (small component peaking near
1.3\,keV).   In PIE, the observed lines are produced at a much cooler
temperature than in CIE.  Therefore, in PIE the bremsstrahlung
continuum is much weaker and there is room for a Compton scattered
continuum (which is represented by the two large continuum
components). \label{ecl_xi}} 

\figcaption[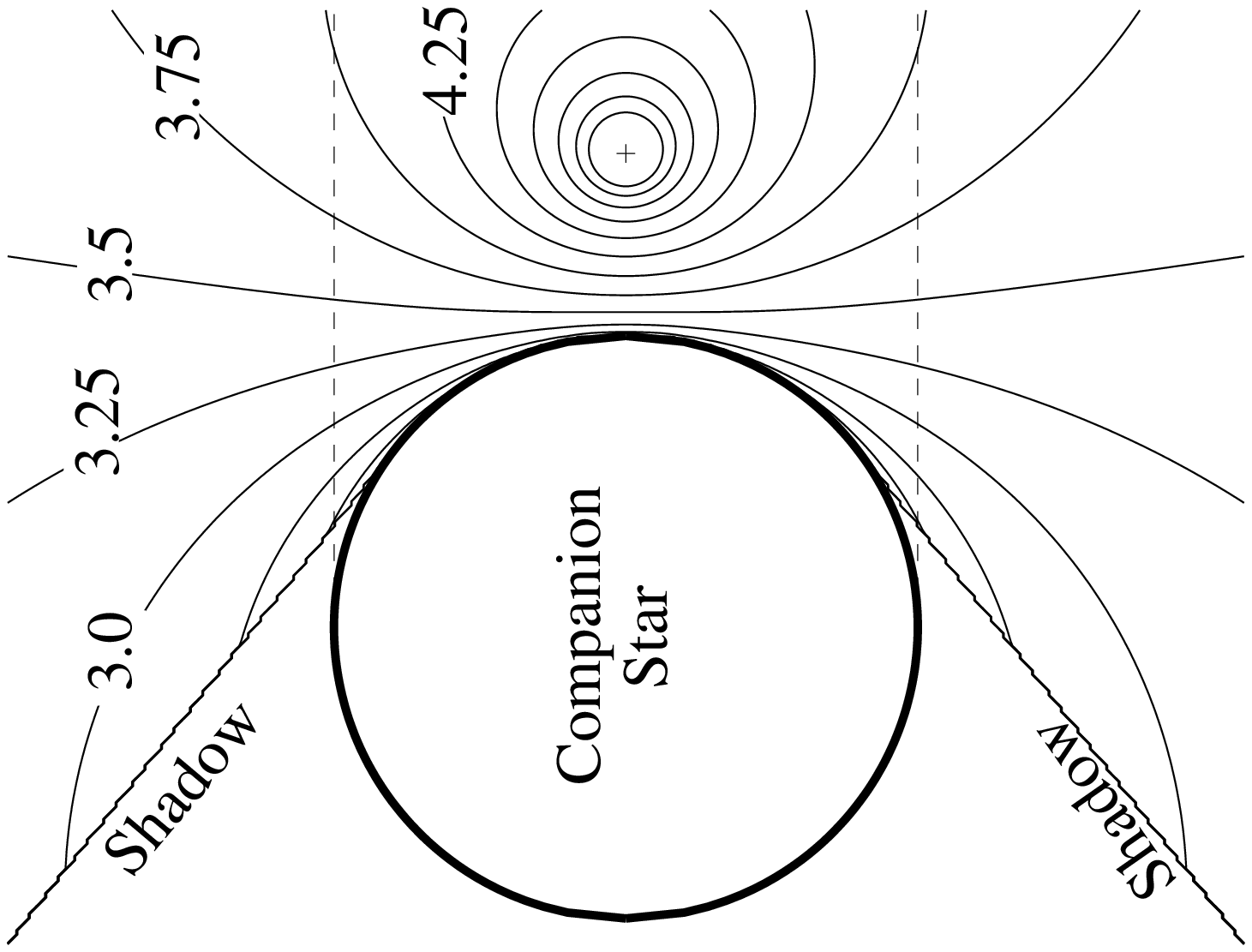]{Contour plot of ionization parameter
distribution for the stellar wind in the orbital plane.  Contours of
constant $\log\xi$ are indicated.  Dashed lines delineate the region
which is not visible when the system is viewed from an orientation
opposite the neutron star.  The position of the neutron star is
indicated by crosshairs. \label{spider}}

\figcaption[f5.ps]{Apparent DEM distributions for the best fit
stellar wind model parameters as viewed at three phases characteristic
of the {\it ASCA\/} observation.  At eclipse center (phase 0) the
regions of highest $\xi$ (near the neutron star) and the regions of
lowest ionization parameter (near the stellar surface) are highly
occulted.  Away from eclipse center the DEM increases at all values of
$\xi$ as more of the wind becomes visible, though the increase is most
dramatic at the highest and lowest values of $\xi$. \label{star_dems}}

\figcaption[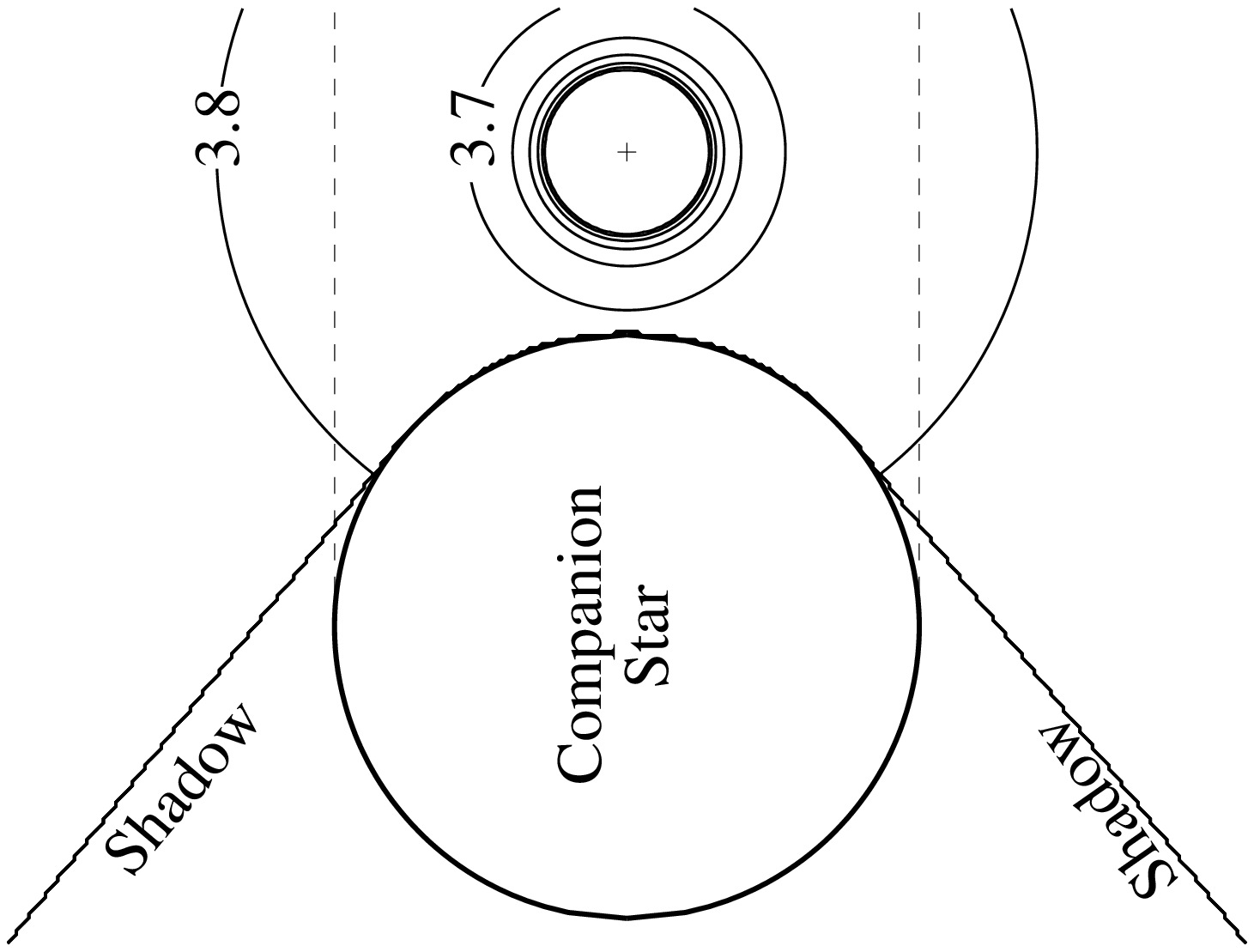]{Contour plot of ionization parameter
distribution for a disk wind with parameters of Table~\ref{fitdata}
except with $R_{\rm in}=3.4R_\sun$, the Roche-lobe radius.  In this
model, the region inside the innermost contour has zero density.  At
large radii, the density varies as $R^{-2}$ and so $\xi$ approaches a
constant value.
\label{spider_disk}} 

\figcaption[f7.ps]{The apparent differential emission measure
for the disk wind for orbital phases characteristic of this
observation.  The model parameters are the same as in
Figure~\ref{spider_disk}.  Most of the emission measure comes from
radii not much larger than $R_{\rm in}$.  Therefore, if $R_{\rm in}$
is much smaller than the companion star, most of the emission measure,
and all of the emission measure at lower values of $\xi$ is occulted
during eclipse.  The apparent emission measure in eclipse is
about one-tenth that outside of eclipse here. \label{disk_dems}}

\figcaption[f8a.ps,f8b.ps,f8c.ps,f8d.ps]{ The best
fit to the \asca spectra with the stellar wind model.  The crosses
represent the data.  The dashed and dotted lines represent the two
absorbed power-law continuum components, the Fe K$\alpha$ line, and
the recombination component and the solid line represents the total
model spectrum. \label{four_fit}}

\figcaption[f9.ps]{ The eclipse phase model spectrum.
The solid line indicates the total flux and the broken lines indicate
the fluxes of the absorbed power-law continua.  Among the features not
labeled are several Fe 3$\rightarrow$2 lines in the 1--2 keV range.
Between \ion{Si}{14} Ly$\beta$ at 2.375 keV and \ion{S}{16} Ly$\alpha$
at 2.621 keV is the \ion{S}{15} 2$\rightarrow$1 complex and higher
order Lyman lines of \ion{Si}{14}.
\label{ecl_unconvolved}} 

\figcaption[f10.ps]{$\chi^2$ vs. inner wind radius for the disk wind
model.  The horizontal lines indicate confidence intervals.  The
vertical line represent the neutron star Roche lobe radius, the
distance from the neutron star to the stellar surface. \label{diskchi}}

\figcaption[f11.ps]{Our best fit stellar wind model line fluxes
($\circ$) and \citet{ebi96} measured line fluxes
($\times$).  Our model line fluxes are always within three sigma
of the fit values.  All fluxes have been corrected for interstellar
absorption ($N_{\rm H1}$).  \label{fig:line_fluxes}} 

\figcaption[f12.ps]{A comparison of the DEM distributions for
Cen~X-3 and Vela~X-1. \label{vc_logdem}}


\clearpage

\includegraphics[width=2.5in,angle=270]{f1.ps}
{\center Figure \ref{ecl_mekal}}

\includegraphics[width=2.5in,angle=270]{f2.ps}
{\center Figure \ref{ecl_mekalv}}

\includegraphics[width=2.5in,angle=270]{f3.ps}
{\center Figure \ref{ecl_xi}}

\includegraphics[width=3.0in,angle=270]{f4.ps}
{\center Figure \ref{spider}}

\includegraphics[width=3.0in,angle=270]{f5.ps}
{\center Figure \ref{star_dems}}

\includegraphics[width=3.0in,angle=270]{f6.ps}
{\center Figure \ref{spider_disk}}

\includegraphics[width=3.0in,angle=270]{f7.ps}
{\center Figure \ref{disk_dems}}

\clearpage

\begin{tabular}{cc}
\includegraphics[angle=270,width=2.5in]{f8a.ps} &
\includegraphics[angle=270,width=2.5in]{f8b.ps} \\
\includegraphics[angle=270,width=2.5in]{f8c.ps} &
\includegraphics[angle=270,width=2.5in]{f8d.ps} \\
\end{tabular}
{\center Figure \ref{four_fit}a,b,c,d}

\clearpage

\includegraphics[angle=270,width=3.0in]{f9.ps}
{\center Figure \ref{ecl_unconvolved}}

\includegraphics[angle=270,width=3.0in]{f10.ps}
{\center Figure \ref{diskchi}}

\includegraphics[width=4.0in,angle=270]{f11.ps}
{\center Figure \ref{fig:line_fluxes}}

\includegraphics[angle=270,width=3.0in]{f12.ps}
{\center Figure \ref{vc_logdem}}

\end{document}